\documentclass[preprint,showpacs,preprintnumbers,amsmath,amssymb,prl]{revtex4}
\usepackage{amsmath}
\usepackage{graphics}
\usepackage{epsf}
\usepackage{dcolumn}
\usepackage{bm}
\begin{document}

\title{Dominant role of local-moment interactions in the magnetism in iron 
pnictides : comparative study of arsenides and antimonides from first-principles
}

\author{Chang-Youn Moon, Se Young Park, and Hyoung Joon Choi}
\email[Email:\ ]{h.j.choi@yonsei.ac.kr}
\affiliation{Department of Physics and IPAP, Yonsei University, Seoul 120-749, Korea}

\date{\today}

\begin{abstract}
The magnetic properties of various iron pnictides are investigated using
first-principles pseudopotential calculations. We consider three different families,
LaFePnO, BaFe$_2$Pn$_2$, and LiFePn with Pn=As and Sb, and find that the Fe local spin
moment and the stability of the 
stripe-type antiferromagnetic phase increases from As to Sb for all 
of the three families, with a partial gap formed at the Fermi energy. 
In the meanwhile, the Fermi-surface nesting is found to be enhanced from Pn=As to Sb for 
LaFePnO, but not for BaFe$_2$Pn$_2$ and LiFePn. These results indicate
that it is not the Fermi surface nesting but the local moment interaction
that determines the stability of the magnetic phase in these materials, 
and that the partial gap is an induced feature by a specific magnetic order.

\end{abstract}

\pacs{71.15.Mb, 71.20.-b, 75.25.+z, 71.18.+y}

\maketitle
The iron pnictide superconductors and their fascinating physical properties have become central issues 
in many fields since their recent discoveries \cite{Kamihara2006,Kamihara2008,Takahashi}.
The prototype materials are REFeAsO with variouof RE (rare earth) elements, 
and the superconducting transition temperature
($T_c$) is as high as 55 K in doped SmFeAsO \cite{Ren2053}.
Other compounds with various types of insulating layers are also
superconducting when doped, such as K-doped BaFe$_2$As$_2$ \cite{0805.4021,0805.4630} and 
SrFe$_2$As$_2$ \cite{0806.1043,0806.1209} with $T_c$ of 38 K, and LiFeAs with $T_c$ of 16 K 
\cite{Pitcher} or 18 K \cite{Wang,Tapp}.
Without doping, these materials exhibit a peculiar magnetic structure of a stripe-type antiferromagnetic (AFM) 
spin configuration coupled to orthorhombic atomic structure, 
and either hole or electron doping destroys the AFM and the 
superconductivity emerges subsequently. 
Hence the magnetism is considered to be closely related to the superconductivity in these materials
\cite{Giovannetti,Singh,Haule,Xu,Mazin}, and the spin-fluctuation-mediated superconductivity 
is assumed in many theoretical works \cite{Mazin2,Kuroki,Korshunov}.

Understanding the nature of magnetism in these materials is thus of crucial importance,
but it still under debate. On one hand, many theoretical \cite{Mazin,Cvetkovic,Yin,Dong}
and experimental \cite{Dong,Lorenz,Cruz,Klauss,Hu} works emphasize on the itinerant nature of the
magnetism of the spin density wave (SDW) type, since the electron and hole Fermi surfaces (FS) are separated by
a commensurate nesting vector in iron pnictides, which is further supported by 
the reduced magnetic moment of about 0.3 $\mu_B$ \cite{Cruz,Klauss} and the energy gap near 
the Fermi energy ($E_F$). On the other hand, there are also interpretations based on the Heisenberg-type
interaction between localized spin moments \cite{Si,Yang2,Yildirim}. 
In this localized-moment picture, the observed stripe-type AFM ordering 
results from the frustrated
spin configuration with the next-nearest-neighbor exchange interaction ($J_2$) larger than half of the
nearest-neighbor (NN) interaction ($J_1$). The itinerant and the local-moment pictures are based on 
different assumptions on the
electron itinerancy, but a more comprehensive mechanism might be discovered by combining the two
pictures  \cite{Wu,Kou}.

Recently, motivated by 
the great success of the As substitution for P in LaFePO on raising $T_c$, hypothetical iron antimonide 
compounds have been studied as candidates for a higher-$T_c$ superconductor by first-principles 
calculations \cite{Moon,Zhang}. In these works, Sb substitution for As is found to modify the FS
nesting and the magnetic stability significantly. Thus, with more variation of compounds including
antimonides, more comprehensive understanding of
the nature of magnetism in iron pnictides would be possible through a systematic comparative 
study dealing with many different types of compounds altogether.

In this study, we present our density-functional pseudopotential calculations of
the electronic and magnetic properties of various iron arsenides and antimonides: LaFePnO, BaFe$_2$Pn$_2$, 
and LiFePn (Pn=As and Sb).
We find that there is no systematic trend of FS nesting feature between arsenides and
antimonides, whereas the stability and the local Fe spin moment
of the magnetic phase 
increase from arsenides to antimonides for all three types of compounds. This finding is consistent with 
Heisenberg-type interaction picture that the local Fe moment is larger for 
antimonides with the enhanced Hund's rule coupling due to their larger lattice constants.
We also find that the FS reconstruction and the subsequent
formation of a partial gap in the density of states (DOS) at $E_F$ can be regarded as
a secondary effect caused by the magnetic ordering of local moments.

Our first-principles calculations are based on the density-functional theory (DFT) within
the generalized gradient approximation (GGA) for the exchange-correlation energy functional
\cite{PBE} and the {\it ab-initio} norm-conserving pseudopotentials as implemented in SIESTA 
code \cite{SIESTA}.
Semicore pseudopotentials are used for Fe, La, and Ba, and
electronic wave functions are expanded with localized pseudoatomic orbitals (double zeta polarization
basis set), with the cutoff energy for real space mesh of 500 Ry. Brillouin zone integration is
performed by Monkhorst-Pack scheme \cite{Monkhorst} with 12 $\times$ 12 $\times$ 6 k-point grid.

First we obtain the optimized cell parameters and atomic coordinates of compounds by total energy
minimization, as listed in Table I. For the non-magnetic (NM) phase, tetragonal structures are obtained  
while the stripe-type AFM phase prefers the orthorhombic structure of the approximate $\sqrt{2} \times
\sqrt{2}$ supercell, in agreement with experiments.
The lowering of the total energy per Fe atom in the stripe-type AFM phase in the optimized orthorhombic 
structure relative to the NM phase
in the optimized tetragonal unit cell is 354 and 706 meV for LaFeAsO and LaFeSbO, 
297 and 745 meV for BaFe$_2$As$_2$ and BaFe$_2$Sb$_2$, and 153 and 523 meV for LiFeAs and LiFeSb, 
respectively. Along with the local magnetic moments on Fe atoms displayed in Table I, this result 
implies the existence of a universal trend that the magnetism is stronger for antimonides than for 
arsenides irrespective of the detailed material properties.

Figure 1 shows the calculated FSs on the $k_z=0$ plane. To facilitate 
the investigation of the nesting feature, the electron and hole surfaces are drawn together in the reduced 
Brillouin zone for the $\sqrt{2} \times \sqrt{2}$ supercell. LaFeSbO shows an enhanced nesting between
the electron and hole surfaces which coincide with each other very isotropically with almost circular 
shapes compared with LaFeAsO \cite{Moon}. For BaFe$_2$Pn$_2$,
the arsenide exhibits a moderate nesting feature, while 
nesting looks poor for the antimonide because hole surfaces, which are present in the arsenide, are missing
so that the electron surfaces have no hole surfaces to couple with nearby. LiFeSb also shows an inefficient
nesting compared with LiFeAs with some hole surfaces missing around the $\Gamma$ point.

The nesting feature can be more quantitatively estimated by evaluating
the Pauli susceptibility $\chi_0({\bf q})$ as a function of the momentum ${\bf q}$ in the static limit with
matrix elements ignored.
The result is displayed in Fig. 2. For LaFePnO, $\chi_0$ is larger for LaFeSbO for entire range of ${\bf q}$,
especially at the nesting vector ${\bf q}=(\pi,\pi)$ where the pronounced peak is located. This peak
indicates the enhanced FS nesting for LaFeSbO, consistently with the FS topology 
in Fig. 1. For BaFe$_2$Pn$_2$, situation is drastically different. Although the susceptibility for 
BaFe$_2$As$_2$ has similar
${\bf q}$ dependence with those for LaFePnO, 
the susceptibility for BaFe$_2$Sb$_2$ is larger only for partial range of ${\bf q}$ with very weak ${\bf q}$ 
dependence and moreover there is no peak at ${\bf q}=(\pi,\pi)$. This feature clearly reflects 
the poor FS
nesting in BaFe$_2$Sb$_2$ due to the lack of hole surfaces, as shown in Fig. 1. Finally, LiFeSb also
has smaller $\chi_0({\bf q})$ than LiFeAs near $(\pi,\pi)$, 
hence LiFeSb has less effective FS nesting at $(\pi,\pi)$ than LiFeAs.

Although many previous studies suggest the itinerant magnetism in iron pnictides that the 
stripe-type AFM is the SDW type driven by the FS nesting, our results are in
contradiction with this picture of magnetism. As we have just
discussed, the FS nesting for ${\bf q}=(\pi,\pi)$, at which the stripe-type AFM occurs,
is more pronounced for LaFeSbO than LaFeAsO, while BaFe$_2$As$_2$ and LiFeAs have more 
effective nesting feature than BaFe$_2$Sb$_2$ and LiFeSb, respectively. 
Thus, there is no universal trend in the FS nesting feature between arsenides and antimonides, which
is in contrast, however, with the result that magnetism is stronger 
for antimonides than the respective arsenides for all three types of iron pnictides, with larger 
energy differences between AFM and NM states and greater Fe local magnetic moments for antimonides.
This implies that the contribution of itinerant electrons to the magnetic energy and moment
is relatively small.

In order to obtain a deeper insight into the nature of magnetism in these
compounds, we consider another type of AFM ordering to examine how the relative
stability and magnetic moments are affected by different AFM ordering. The additional AFM ordering considered
is a `checkerboard' type AFM ordering in which the four NN Fe atoms 
have the opposite spin direction to the Fe atom which they surround. This AFM ordering 
is denoted by AFM1 in this paper, and the stripe-type AFM ordering
by AFM2. In Table II, the relative energy of each AFM type 
and the magnetic moment on a Fe atom are listed for all the six compounds. For each compound,
atomic structures optimized in the NM phase are used
for all magnetic phases to see purely electronic
contribution to the total energy differences among magnetic phases without structural relaxation 
effects. 

As shown in Table II, AFM1 is more stable than NM phase for all of the compounds, and
the stability and the Fe local magnetic moment are larger for the antimonides than their respective 
arsenides. Since the AFM1 ordering is surely not related to the FS nesting, 
there should be a mechanism other than the simple itinerant magnetism to explain the stability of
AFM1 and its enhancement in antimonides. 
Furthermore, we find energetic stability of AFM2 relative to AFM1 phases and magnetic moment 
in AFM2 phase are enhanced
in all antimonides compared with respective arsenides, as shown in Table II. This is again in 
contradiction with
FS nesting features related to the itinerant magnetism. Therefore, the Heisenberg-type
magnetic interaction naturally arises as more appropriate description for the magnetism in
these materials. As the lattice parameters are larger for antimonides than their corresponding
arsenides, the Fe 3$d$ orbitals are more localized as is evident from the reduced band width
around $E_F$ \cite{Moon}. Thus the Hund's rule coupling becomes stronger
and the local magnetic moment is larger for antimonides, as is in Table II. The generally larger
Fe magnetic moments can explain the enhanced stability of AFM1 with respect to the NM phase, 
and AFM2 with respect to AFM1, for antimonides compared with arsenides within the Heisenberg 
interaction with $J_2 > J_1/2$ \cite{Si,Yang2,Yildirim}.

In the meanwhile, there is clear difference in DOS between AFM2 and other 
phases calculated with the same structural parameters optimized for the NM phase
for each compound, as displayed in Fig. 3.
The NM phase has a finite DOS 
at $E_F$, and AFM1 magnetic ordering does not reduce the DOS at $E_F$, while it is 
greatly reduced for the AFM2 ordering. This feature indicates that the AFM2 phase
involves the ordering-induced FS reconstruction by the coupling between the 
electron and hole surfaces, in contrast to the AFM1 phase where only the local magnetic
interaction is involved. Our result qualitatively agrees with
the recently suggested model \cite{Kou} in which the itinerant electrons couple to the
local magnetic moments which are AFM ordered. Even in the case of BaFe$_2$Sb$_2$,
where the FS nesting is very ineffective as in Figs. 1 and 2, the AFM2 ordering produces
the strong perturbing potential for the electron and hole bands to be hybridized,
resulting in the partial gap in DOS at $E_F$, as shown in Fig. 3 (d). Other compounds
exhibit similar feature in DOS at $E_F$ among different magnetic phases, indicating
that the presence of partial gap is not sensitive to the detailed FS nesting characteristics
as it is an induced feature by coupling to more robust underlying magnetism of the
local moment interaction.

In summary, we investigate the magnetic properties of known and hypothetical iron pnictides 
by the total energy calculations.
We find that our calculated FS nesting feature in the NM phase is not consistent 
with the trend of the magnetic
stability that the AFM phases are more stable 
in antimonides than in arsenides. 
Heisenberg-type local moment interaction is more appropriate to explain our 
results when we consider the larger Fe spin moment found in antimonides.
Thus our results indicate that experimentally observed stripe-type AFM in iron
pnictides is mainly driven by local moment interaction, while SDW of the itinerant electrons
and the partial gap at $E_F$ emerge as an induced order by coupling to the local moments.

\begin{acknowledgments}
This work was supported by the KRF (KRF-2007-314-C00075) and 
by the KOSEF Grant No. R01-2007-000-20922-0. Computational resources 
have been provided by KISTI Supercomputing Center (KSC-2008-S02-0004). 
\end{acknowledgments}


\begin{thebibliography}{35}
\expandafter\ifx\csname natexlab\endcsname\relax\def\natexlab#1{#1}\fi
\expandafter\ifx\csname bibnamefont\endcsname\relax
  \def\bibnamefont#1{#1}\fi
\expandafter\ifx\csname bibfnamefont\endcsname\relax
  \def\bibfnamefont#1{#1}\fi
\expandafter\ifx\csname citenamefont\endcsname\relax
  \def\citenamefont#1{#1}\fi
\expandafter\ifx\csname url\endcsname\relax
  \def\url#1{\texttt{#1}}\fi
\expandafter\ifx\csname urlprefix\endcsname\relax\def\urlprefix{URL }\fi
\providecommand{\bibinfo}[2]{#2}
\providecommand{\eprint}[2][]{\url{#2}}

\bibitem[{Kam({\natexlab{a}})}]{Kamihara2006}
\bibinfo{note}{Y. Kamihara {\it et al.}, J. Am. Chem. Soc. {\bf 128}, 10012, (2006).}

\bibitem[{Kam({\natexlab{b}})}]{Kamihara2008}
\bibinfo{note}{Y. Kamihara, T. Watanabe, M. Hirano, and H. Hosono, J. Am. Chem.
  Soc. {\bf 130}, 3296, (2008).}

\bibitem[{Tak()}]{Takahashi}
\bibinfo{note}{H. Takahashi {\it et al.}, Nature (London) {\bf 453}, 376 (2008).}

\bibitem[{Ren({\natexlab{a}})}]{Ren2053}
\bibinfo{note}{Z.-A. Ren {\it et al.}, Chin. Phys. Lett. {\bf 25}, 2215 (2008).}

\bibitem[{080({\natexlab{a}})}]{0805.4021}
\bibinfo{note}{M. Rotter {\it et al.}, Phys. Rev. B {\bf 78}, 020503(R) (2008).}

\bibitem[{080({\natexlab{b}})}]{0805.4630}
\bibinfo{note}{M. Rotter, M. Tegel, and D. Johrendt, Phys. Rev. Lett. {\bf 101}, 107006 (2008).}

\bibitem[{080({\natexlab{c}})}]{0806.1043}
\bibinfo{note}{C. Krellner {\it et al.}, Phys. Rev. B {\bf 78}, 100504(R), 2008.}

\bibitem[{080({\natexlab{d}})}]{0806.1209}
\bibinfo{note}{G. F. Chen {\it et al.}, Chin. Phys. Lett. {\bf 25}, 3403 (2008).}

\bibitem[{Pit()}]{Pitcher}
\bibinfo{note}{M. J. Pitcher {\it et al.}, Chem. Commun. 5918 (2008).}

\bibitem[{Wan()}]{Wang}
\bibinfo{note}{ X. C. Wang {\it et al.}, Solid State Commun. {\bf 148}, 538 (2008).}

\bibitem[{Tap()}]{Tapp}
\bibinfo{note}{J. H. Tapp {\it et al.}, Phys. Rev. B {\bf 78}, 060505(R) (2008).}

\bibitem[{Gio()}]{Giovannetti}
\bibinfo{note}{G. Giovannetti, S. Kumar, and J. van den Brink,
  arXiv:0804.0866v2 [Phys. Rev. B (in press)].}

\bibitem[{Sin()}]{Singh}
\bibinfo{note}{D. J. Singh and M.-H. Du, Phys. Rev. Lett. {\bf 100}, 237003
  (2008).}

\bibitem[{Hau()}]{Haule}
\bibinfo{note}{K. Haule, J. H. Shim, and G. Kotliar, Phys. Rev. Lett. {\bf
  100}, 226402 (2008).}

\bibitem[{Xu()}]{Xu}
\bibinfo{note}{G. Xu {\it et al.}, Europhys. Lett. {\bf 82}, 67002 (2008).}

\bibitem[{Maz()}]{Mazin}
\bibinfo{note}{I. I. Mazin {\it et al.}, arXiv:0806.1869v2.}


\bibitem[{Maz()}]{Mazin2}
\bibinfo{note}{I. I. Mazin, D. J. Singh, M. D. Johannes, and M. H. Du, Phys. Rev. Lett. {\bf 101}, 
057003 (2008).}

\bibitem[{Kur()}]{Kuroki}
\bibinfo{note}{K. Kuroki {\it et al.}, Phys. Rev. Lett. {\bf 101}, 087004 (2008).}

\bibitem[{Kor()}]{Korshunov}
\bibinfo{note}{M. M. Korshunov and I. Eremin, Phys. Rev. B {\bf 78}, 140509(R) (2008).}

\bibitem[{Cve()}]{Cvetkovic}
\bibinfo{note}{V. Cvetkovic and Z. Tesanovic, arXiv:0804.4678v3.}

\bibitem[{Yin()}]{Yin}
\bibinfo{note}{Z. P. Yin {\it et al.}, Phys. Rev. Lett. {\bf 101}, 047001 (2008).}

\bibitem[{Don()}]{Dong}
\bibinfo{note}{J. Dong {\it et al.}, Europhys. Lett. {\bf 83}, 27006 (2008).}

\bibitem[{Lor()}]{Lorenz}
\bibinfo{note}{B. Lorenz {\it et al.}, Phys. Rev. B {\bf 78}, 012505 (2008).}

\bibitem[{Cru()}]{Cruz}
\bibinfo{note}{C. de la Cruz {\it et al.}, Nature (London) {\bf 453}, 899 (2008).}

\bibitem[{Kla()}]{Klauss}
\bibinfo{note}{H. -H. Klauss {\it et al.}, Phys. Rev. Lett. {\bf 101}, 077005 (2008).}

\bibitem[{Hu()}]{Hu}
\bibinfo{note}{W. Z. Hu {\it et al.}, arXiv:0806.2652v4.}

\bibitem[{Si()}]{Si}
\bibinfo{note}{Q. Si and E. Abrahams, Phys. Rev. Lett. {\bf 101}, 076401 (2008).}

\bibitem[{Yan()}]{Yang2}
\bibinfo{note}{L. X. Yang {\it et al.}, arXiv:0806.2627v2.}

\bibitem[{Yil()}]{Yildirim}
\bibinfo{note}{T. Yildirim, Phys. Rev. Lett. {\bf 101}, 057010 (2008).}

\bibitem[{Wu()}]{Wu}
\bibinfo{note}{J. Wu, P. Phillips, and A. H. Castro Neto, Phys. Rev. Lett. {\bf 101}, 126401 (2008).}

\bibitem[{Kou()}]{Kou}
\bibinfo{note}{S.-P. Kou, T. Li, and Z.-Y. Weng, arXiv:0811.4111v3.}

\bibitem[{Moo()}]{Moon}
\bibinfo{note}{C.-Y. Moon, S. Y. Park, and H. J. Choi, Phys. Rev. B {\bf 78}, 212507 (2008).}

\bibitem[{Kur()}]{Zhang}
\bibinfo{note}{L. Zhang, A. Subedi, D. J. Singh, and M. H. Du, arXiv:0808.2653v3.}

\bibitem[{PBE()}]{PBE}
\bibinfo{note}{J. P. Perdew, K. Burke, and M. Ernzerhof, Phys. Rev. Lett. {\bf
  77}, 3865 (1996).}

\bibitem[{SIE()}]{SIESTA}
\bibinfo{note}{D. Sanchez-Portal, P. Ordejon, E. Artacho, and J. M. Soler, 
Int. J. Quantum Chem. {\bf 65}, 453 (1997).}

\bibitem[{Mon()}]{Monkhorst}
\bibinfo{note}{H. J. Monkhorst and J. D. Pack, Phys. Rev. B {\bf 13}, 5188
  (1976).}

\bibitem[{Mon()}]{Huang}
\bibinfo{note}{Q. Huang {\it et al.}, Phys. Rev. Lett. {\bf 101}, 257003 (2008).}

\end{thebibliography}

\newpage

\begin{table}
\caption{Calculated structure parameters, DOS at $E_f$ ($N(E_f)$), and Fe magnetic moment ($m$) of 
LaFeAsO (LFAO), LaFeSbO (LFSO), BaFe$_2$As$_2$ (BFA), BaFe$_2$Sb$_2$ (BFS),
LiFeAs (LFA), and LiFeSb (LFA). Both for NM and SDW phases, 
$c$ lattice parameter is taken as the distance of two adjacent Fe layers 
for easier comparison. $z_1$ represents the $z$ coordinate
of La, Ba, or Li, and $z_2$ represents that of As or Sb. Iron atoms are located 
at $z = 0.5$ along the c-axis.
}
\begin{center}
\begin{tabular}{lcccccc}
\hline\hline
\multicolumn{7}{c}{NM (tetragonal)}  \\
          & ~$a$ (\AA) ~ & ~$b$ (\AA) ~ & ~$c$ (\AA)~ & ~$z_1$~ &~$z_2$~ & $N(E_f)$\\
\hline
 LFAO        & 3.999 & 3.999  & 8.706  & 0.145 &  0.640  &  1.7 \\
 LFSO        & 4.106 & 4.106  & 9.311  & 0.130 &  0.659  &  2.9 \\
 BFA         & 3.935 & 3.935  & 6.314  & 0  &  0.696  &  1.9 \\
 BFS         & 4.324 & 4.324  & 6.315  & 0 &  0.708  &  1.8 \\
 LFA         & 3.767 & 3.767  & 5.967  & 0.173 &  0.734  &  2.1 \\
 LFS         & 3.995 & 3.995  & 6.266  & 0.211 &  0.756  &  2.6 \\
\hline
\multicolumn{7}{c}{SDW (orthorhombic) }  \\
         & ~$a$ (\AA) ~ & ~$b$ (\AA) ~ & ~$c$ (\AA)~ & ~$z_1$~ &~$z_2$~ & $m(\mu_B)$\\
\hline
 LFAO        & 5.780 & 5.693 & 8.875 & 0.139 &  0.654  &  2.83 \\
 LFSO        & 5.955 & 5.844 & 9.542 & 0.124 &  0.673  &  3.13 \\
 BFA         & 5.756 & 5.590 & 6.520 & 0 &  0.712  &  2.78 \\
 BFS         & 6.231 & 5.937 & 7.246 & 0 &  0.722  &  3.22 \\
 LFA         & 5.482 & 5.285 & 6.190 & 0.171 &  0.745  &  2.54 \\
 LFS         & 5.830 & 5.593 & 6.528 & 0.199 &  0.768  &  2.95 \\
\hline\hline
\end{tabular}
\end{center}
\label{table I}
\end{table}

\begin{table}
\caption{Stability of magnetic phases and Fe magnetic moments $m$ in $\mu_B$ for iron pnictides. 
For each compound, calculations are done in the optimized structure for the NM phase.
$E_1$ is the energy of AFM1 relative to the NM phases and $E_2$ is the energy of AFM2
relative to the AFM1 phases, in meV per Fe atom.
}
\begin{center}
\begin{tabular}{lcccc}
\hline\hline
Compound        & $E_1$  & $m$(AFM1) &$E_2$ & $m$(AFM2) \\
\hline
LaFeAsO         & -123  &  2.23  &  -109  &  2.35 \\
LaFeSbO         & -387  &  2.88  &  -136  &  2.83 \\
BaFe$_2$As$_2$  & -108  &  2.09  &  -64   &  2.20 \\
BaFe$_2$Sb$_2$  & -426  &  2.80  &  -75   &  2.78 \\
LiFeAs          & -45   &  1.83  &  -99   &  1.96 \\
LiFeAs          & -269  &  2.54  &  -118  &  2.63 \\
\hline\hline
\end{tabular}
\end{center}
\label{table II}
\end{table}

\newpage

\begin{figure}
\caption{(Color online). Calculated FS of iron pnictides in the NM phase, drawn in
the Brillouin zone of the $\sqrt{2} \times \sqrt{2}$ supercell (dashed lines).
Hole pockets are represented in blue (dark gray), and electron pockets are in red (gray). 
In (c) and (d), the conventional simple tetragonal unit cell is used rather than the primitive
body-centered tetragonal unit cell, for easier comparison with other compounds.
}
\label{fig1}
\end{figure}

\begin{figure}
\caption{(Color online). Pauli susceptibility $\chi_0({\bf q})$ for (a) LaFePnO, (b)
BaFe$_2$Pn$_2$, and (c) LiFePn, normalized by $\chi_0({\bf q}=(0,0))$ of the arsenide
for each type of compounds.
}
\label{fig2}
\end{figure}

\begin{figure}
\caption{(Color online). DOSs of (a) LaFeAsO, (b) LaFeSbO, (c) BaFe$_2$As$_2$, (d) BaFe$_2$Sb$_2$,
(e) LiFeAs, and (f) LiFeSb calculated in the NM (black), AFM1 (red), and AFM2 (blue) phases.
}
\label{fig3}
\end{figure}

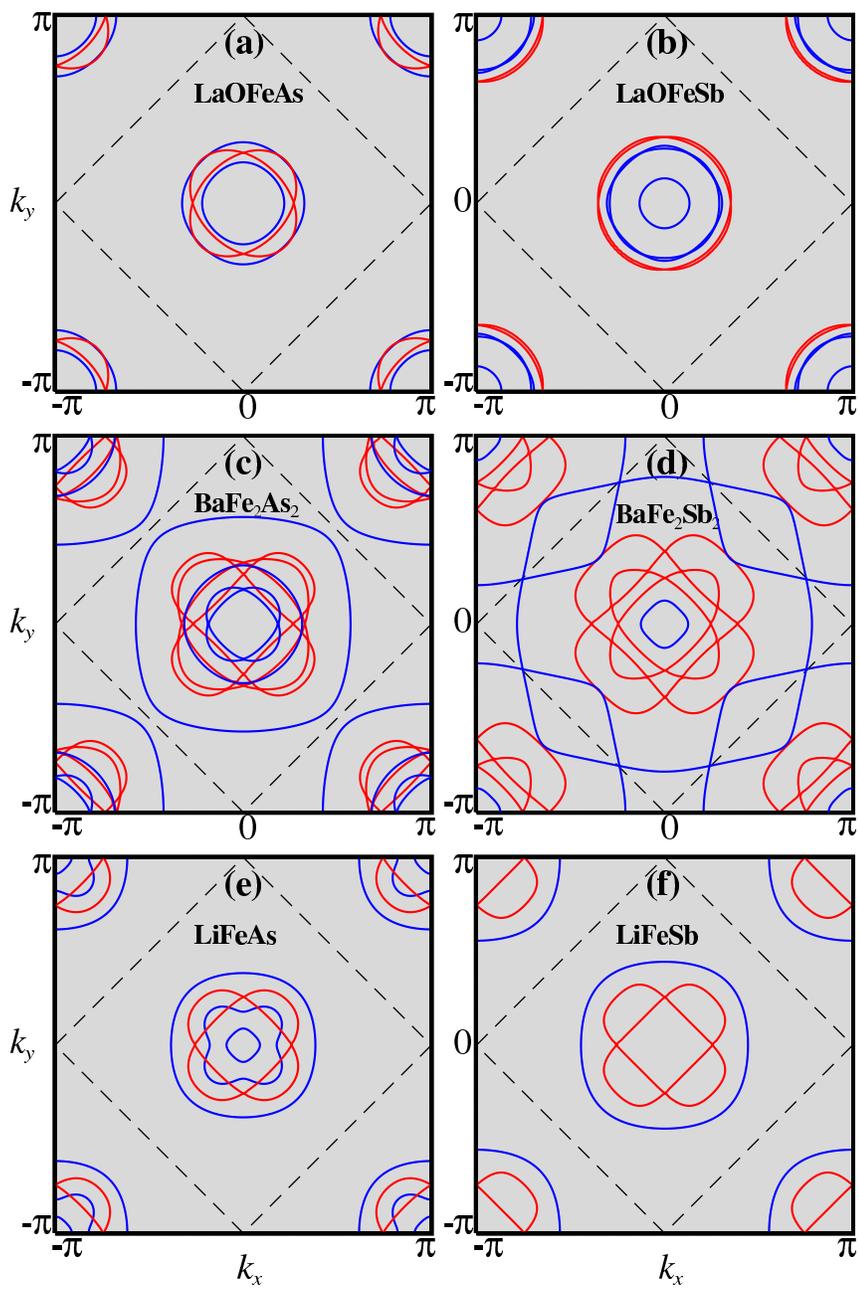
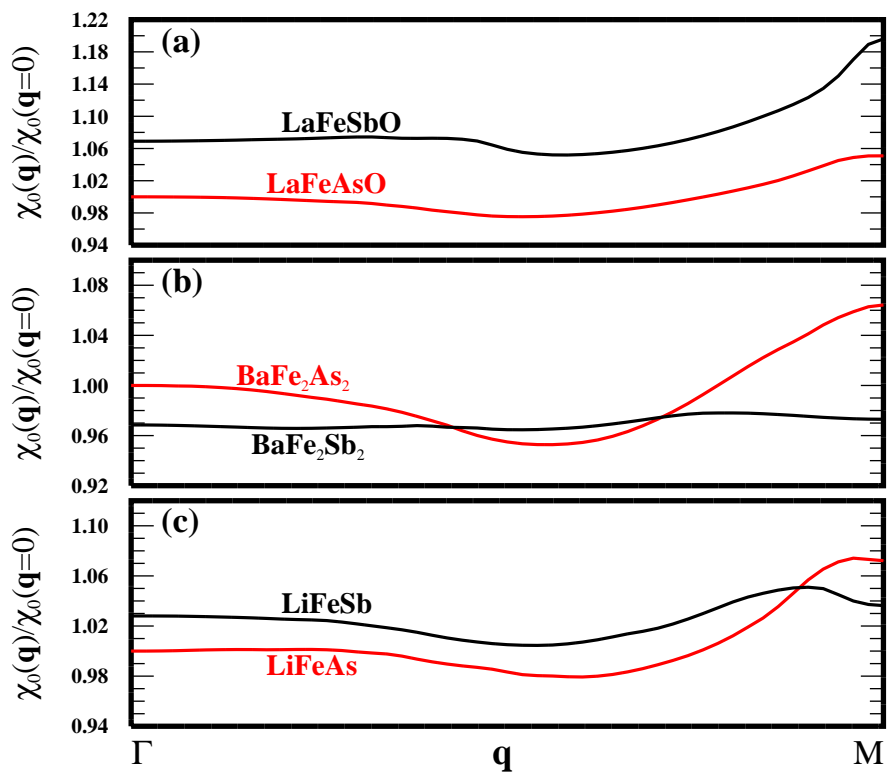
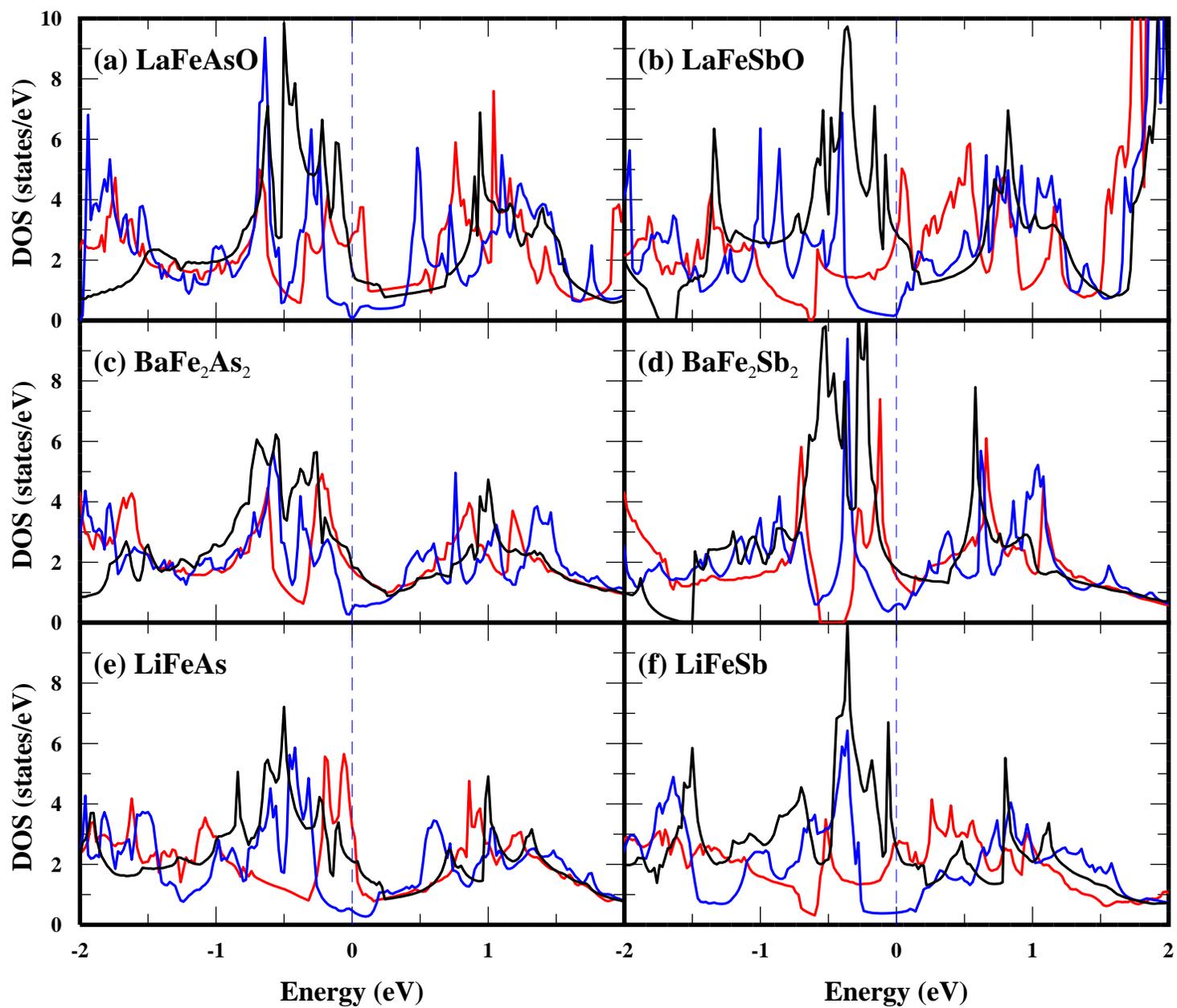
\end{document}